\documentclass[%
 reprint,
 amsmath,amssymb,
 aps,
]{revtex4-2}

\usepackage{graphicx}
\usepackage{dcolumn}
\usepackage{bm}
\usepackage{hyperref}

\newcommand{\km}{k_\mathrm{m}}

\newcommand{\ko}{k_\mathrm{o}}

\newcommand{\om}{\omega_{\rm m}}
\newcommand{\oo}{\omega_{\rm o}}
\newcommand{\E}{\boldsymbol{E}}
\newcommand{\Ef}{\tilde{\boldsymbol{E}}_\mathrm{f}}
\newcommand{\Eb}{\tilde{\boldsymbol{E}}_\mathrm{b}}
\newcommand{\Q}{\boldsymbol{u}}
\newcommand{\Qf}{\tilde{\boldsymbol{u}}_\mathrm{f}}
\newcommand{\Qb}{\tilde{\boldsymbol{u}}_\mathrm{b}}
\newcommand{\kmf}{k_\mathrm{m,f}}
\newcommand{\kmb}{k_\mathrm{m,b}}
\newcommand{\kof}{k_\mathrm{o,f}}
\newcommand{\kob}{k_\mathrm{o,b}}
\renewcommand{\d}{\mathrm{d}}
\renewcommand{\r}{\boldsymbol{r}}

\begin{document}

\preprint{APS/123-QED}

\title{Clamped and sideband-resolved silicon optomechanical crystals}

\author{Johan Kolvik}
 \altaffiliation{These authors contributed equally to this work.}
\author{Paul Burger}%
 \altaffiliation{These authors contributed equally to this work.}
 \author{Joey Frey}
\author{Rapha\"{e}l Van Laer}%
 \email{raphael.van.laer@chalmers.se}
\affiliation{Department of Microtechnology and Nanoscience (MC2), Chalmers University of Technology.}

\date{\today}

\begin{abstract}
Optomechanical crystals (OMCs) are a promising and versatile platform for transduction between mechanical and optical fields. However, the release from the substrate used in conventional suspended OMCs also prevents heat-carrying noise phonons from rapidly leaking away. Thermal anchoring may be improved by attaching the OMCs directly to the substrate. Previous work towards such \textit{clamped}, i.e. non-suspended, OMCs suffers from weak interaction rates and insufficient lifetimes. Here, we present a new class of clamped OMCs realizing -- for the first time -- optomechanical interactions in the resolved-sideband regime required for quantum transduction. Our approach leverages high-wavevector mechanical modes outside the continuum. We observe a record zero-point optomechanical coupling rate of $g_0/(2\pi) \approx 0.50$ MHz along with a sevenfold improvement in the single-photon cooperativity of clamped OMCs. Our devices operate at frequencies commonly used in superconducting qubits. This opens a new avenue using clamped OMCs in both classical and quantum communications, sensing, and computation through scalable mechanical circuitry that couples strongly to light.
\end{abstract}

\maketitle

The field of optomechanics receives great interest for applications such as sensing and microwave-to-optics transduction \cite{mirhosseini_superconducting_2020, jiang_optically_2022,safavi-naeini_controlling_2019,weaver2022integrated}. A leading class of optomechanical devices is the optomechanical crystal (OMC) \cite{chan_optimized_2012}. In state-of-the-art suspended OMCs, the low-wavevector GHz mechanical modes are confined partly by suspending the device layer. Along with engineered bandgaps, this eliminates mechanical leakage into the substrate \cite{chan_optimized_2012}. However, suspension also comes at the cost of losing a channel through which heat-carrying noise phonons created by optical absorption can dissipate \cite{meenehan_silicon_2014}. 

Confining coherent GHz phonons while letting heat-carrying noise phonons leak away fast is an understudied and challenging problem. One approach is to laterally connect a suspended OMC region with the rest of the device layer. While these two-dimensional OMCs have led to impressive results, they also require in-plane bandgaps and the associated fine-tuning of geometry \cite{ren_two-dimensional_2020}. Another approach to provide thermal anchoring is to attach the OMC directly to a substrate. We call this unconventional class of OMCs \textit{clamped}. Besides reducing fabrication complexity and improving thermal anchoring, clamped devices could ease co-integration between phononic, electronic, and photonic devices, all commonly fabricated in the silicon-on-insulator (SOI) platform. Efforts along these lines have been made by using e.g. bound states in the continuum \cite{liu_optomechanical_2022} and geometrical softening \cite{zhang_subwavelength_2022, sarabalis_release-free_2017}. These approaches have not yet been able to compete with the conventional suspended systems because of weaker interactions and shorter coherence times. 

Here, we propose and demonstrate a new class of clamped, i.e. unreleased, OMCs. In our SOI-based OMCs the optomechanical three-wave-mixing interaction takes places between two counter-propagating optical modes and a high-wavevector mechanical mode. We demonstrate the first clamped OMCs with mechanical frequencies exceeding their optical loss rates. This resolved-sideband condition is essential for low-noise quantum transduction between optical and mechanical fields \cite{aspelmeyer_cavity_2014}. Our new clamped OMCs have record zero-point optomechanical coupling rates of $g_0/(2\pi)=0.50 \pm 0.01 \text{ MHz}$ at mechanical frequencies of $\omega_{\rm m}/(2\pi)=5.4\text{ GHz}$. In addition, our clamped OMCs can have a thermal contact area exceeding that of their suspended counterparts. Our results provide a new path for unreleased optomechanical structures to become a competitive platform for both classical and quantum optomechanical circuits. In the following, we first outline the design process of the clamped OMCs and move on to show finite-element simulations of both photonic-phononic crystal waveguides and resonators. Finally, we present fabrication and measurement results of our devices at room temperature.

\begin{figure}[ht!]
\centering\includegraphics{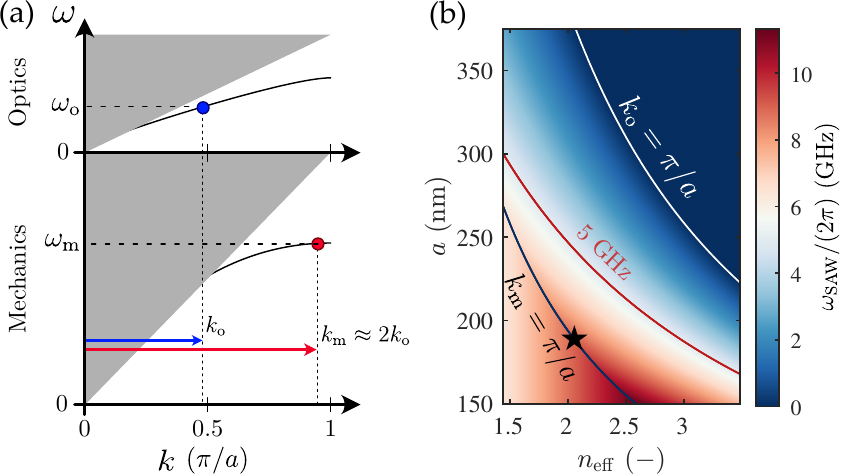}
\caption{\textbf{(a)} Phase-matching diagram for clamped and counter-propagating optomechanical interactions. The optical mode (blue) with frequency $\oo$ couples to a mechanical mode at $\om$. Counter-propagating optical modes interact with a phase-protected mechanical mode at $\km\approx 2\ko$ (red) whose wavevector lies outside the continuum of mechanical modes in the substrate. \textbf{(b)} SAW frequencies in silicon dioxide at the mechanical wavevector $\km = 2\ko$. The optical pump has a vacuum wavelength $\lambda_0 = 1550$ nm and effective index $n_\mathrm{eff}$ and interacts with mechanics in an OMC with unit cell with period $a$. Phase-protection from both optical and mechanical continua is achieved for all $(n_\mathrm{eff},a)$ where $n_\mathrm{eff} > n_\mathrm{cladding}=1.45$ and $\om <  \omega_\mathrm{SAW}.$ The operating point for the presented OMC's defect unit cell is marked by a star.}
\label{fig1}
\end{figure}

\label{sec:design}
\textbf{Design.} A key challenge in designing clamped OMCs is avoiding excessive mechanical leakage into the substrate while keeping a high optomechanical interaction rate. In contrast to its suspended counterpart, a clamped OMC supports a continuum of states for \textit{both} optical and mechanical waves. Here, we limit leakage from the OMC mechanical mode to substrate bulk- and surface-acoustic-waves (SAWs) by exploiting large mechanical wavevectors. This enables the mechanical mode to be phase-protected from the acoustic continuum, similar to optical waveguides based on total internal reflection. Realizing strong three-wave-mixing optomechanical interactions with such high-wavevector mechanical modes requires that $\km \approx 2\ko$ where $\km$ $(\ko)$ is the mechanical (optical) wavevector. This phase-matching condition is familiar from the realm of counter-propagating Brillouin interactions \cite{eggleton_brillouin_2022}. We illustrate the principles of phase-matching for our clamped OMCs in Fig.\ref{fig1}a. 

For a mechanical mode with frequency $\om$ to fall below the continuum, the mechanical wavevector must satisfy $\km >\om /v_\mathrm{SAW}$, where $v_\mathrm{SAW}$ is the substrate SAW phase-velocity. In addition, the first Brillouin zone associated with the crystal period $a$ sets an upper bound to the set of unique wavevectors. This defines an operating window where both mechanical and optical modes are outside their continua while their three-wave-mixing is phase-matched ($\km \approx 2\ko$). To visualize this operating window, we plot SAW frequencies for different operating points in Fig.\ref{fig1}b. At a given wavevector $\km$, there is a maximum frequency the OMC mechanical mode can have before entering the continuum. This upper bound for guided mechanical mode frequencies is calculated as $\omega_\mathrm{SAW} = \km v_\mathrm{SAW}$, where $\km = 2\ko$. We use optical wavevectors for a pump with effective index $n_\mathrm{eff}$ at a vacuum wavelength $\lambda_0 = 1550$ nm. In addition, we use $v_\mathrm{SAW} \approx 3400\ \mathrm{m/s}$ for silicon dioxide \cite{auld_acoustic_1973}. 
Clamped and phase-matched OMC operation with low mechanical and optical radiation losses into the substrate is accessible for all $(n_\mathrm{eff},a)$ such that $n_\mathrm{eff} > n_\mathrm{cladding} = 1.45$ while also $\om < \omega_\mathrm{SAW}$. This analysis encourages exploration of unit cells with smaller periods $a$ as they provide larger operating windows.

\begin{figure}[t!]
\centering\includegraphics{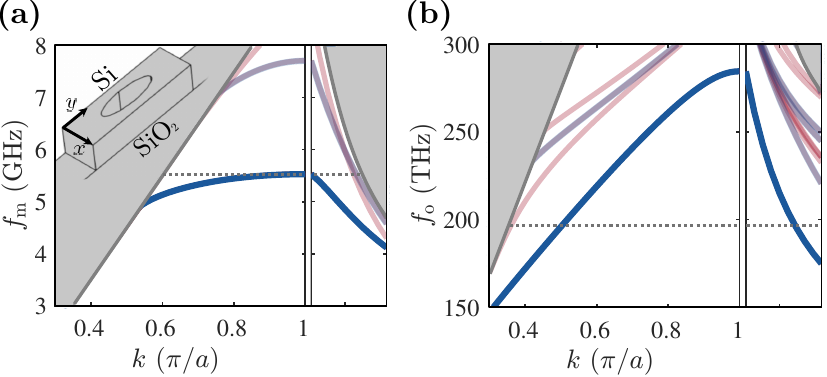}   
\caption{Mechanical (a) and optical (b) band diagram of the OMC defect unit cell. The left inset in (a) shows the unit cell with elliptic hole. The right insets in (a) and (b) show the X-point frequencies as a function of the perturbation from defect to mirror cell. The shaded area denotes the realm of continuum modes. Blue lines denote modes with symmetry with respect to the $xz$ plane, red lines denote modes of other symmetries. The modes of interest are solid. The horizontal dashed lines indicate the approximate OMC mode frequencies at the operating point.}
\label{fig2}
\end{figure}

\begin{figure}[ht!]
\centering\includegraphics[width=1\linewidth]{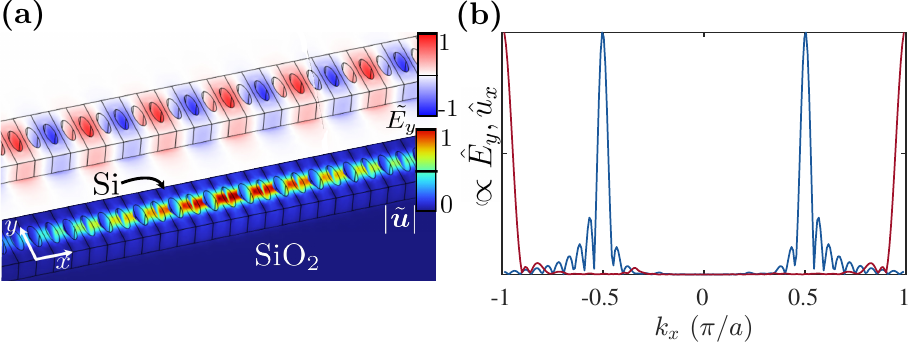}
\caption{(a) Optical (top) and mechanical (bottom) mode profiles of the clamped OMC with $\Tilde{E}_y$ the normalized electric field of the optical mode and $\Tilde{\mathbf{u}}$ the normalized mechanical displacement. (b) Fourier transform of the optical field $E_y$ (blue) and mechanical field $u_x$ (red).}
\label{fig3}
\end{figure}

Investigating unit cells with periods around $a = 188$ nm and widths around $w=643$ nm, we find a guided $X$-point mechanical mode around $\om/(2\pi) = 5.4$ GHz. At this mechanical wavevector, we calculate the SAW frequency to be $\omega_\mathrm{SAW}/(2\pi)=10.3$ GHz. This places our modes firmly outside both the mechanical and optical continua. The mechanical mode profile resembles the ``pinch mode'' first reported in \cite{eichenfield_optomechanical_2009} for suspended OMCs (Fig.\ref{fig3}a). The mechanical motion is primarily longitudinal along the OMC, in contrast to recently explored modes \cite{ma_semiconductor--diamond_2023}. The mode is guided despite the silicon device layer having faster speed of sound than the substrate. We attribute this to the unit cell's relatively large surface-to-volume ratio. This is known to reduce the effective stiffness of the structure in an effect known as geometrical softening \cite{sarabalis_release-free_2017,safavi-naeini_controlling_2019,lagasse_higherorder_1973}. We design a defect unit cell based on this mechanical mode and calculate mechanical and optical dispersion diagrams for a photonic-phononic crystal waveguide with periodic symmetry (Fig.\ref{fig2}). We choose the parameters of the unit cell such that the waveguide supports a C-band optical mode at half the mechanical wavevector. Considering counter-propagating optomechanical interactions, we calculate a unit cell zero-point optomechanical coupling rate of $g_{\rm 0, uc}/(2\pi)=4.2\ \text{MHz}$. This large coupling is mostly mediated by the moving boundary effect and is comparable with those of conventional suspended OMC defect cells \cite{chan_optimized_2012}. We give further details on the implications of high-wavevector mechanics in standing-wave OMCs in appendix.

To confine the waveguide mode described above in a standing-wave OMC, we design a second type of unit cell: the mirror cell. By increasing its period compared to the defect cell by about a factor two to $a = 375$ nm, the optical and mechanical bands are pulled below the localized mode frequencies, opening a quasi-bandgap (Fig.\ref{fig2}). The OMC is thus assembled by transforming the defect cell in the center into the mirror cell at the cavity perimeters \cite{safavi-naeini_design_2010,chan_optimized_2012}. In this case, the mechanical quasi-bandgap only exists for the first half of the perturbation before the mirror cell becomes a host to the continuum modes. Yet, a mechanical quasi-bandgap lasting for only a few unit cells suffices to reflect the vast majority of the mechanical field.

A consequence of the optical mode not being at the $X$-point is that it is not the cavity's fundamental- but a higher order mode. Additionally, the optical mode decays slower in the mirror transition region compared to $X$- or $\Gamma$-point modes. Both of these effects reduce the spatial overlap between optical and mechanical modes. We keep most of the defect unit cell's interaction strength in the full OMC by adding additional defect unit cells before starting the mirror region. Indeed, the field profiles of mechanics and optics approach the defect unit cell's fields for a clamped OMC as the number of defect unit cells $N$ increases. While this reduces the zero-point coupling rates as $1/\sqrt{N}$, the increased interface area between a longer clamped OMC and the substrate is also expected to improve thermal anchoring.

At this stage, we optimize our design using a Nelder-Mead algorithm. For an optimized clamped OMC with $N=31$ defect cells, we simulate an optical mode with $\omega_{\rm o}/(2\pi)=195\ \text{THz}$ and a mechanical mode with $\omega_{\rm m}/(2\pi)=5.64\ \text{GHz}$ with radiation-limited quality factors $Q_{\rm o}=1.3\cdot 10^{6}$ and $Q_{\rm m}=3.6\cdot 10^{5}$ respectively (Fig.\ref{fig3}a). Fourier transformation of the cavity fields indicates that the phase-matching condition $\km \approx 2\ko$ is met for counter-propagating optomechanical interaction (Fig.\ref{fig3}b). The simulated zero-point optomechanical coupling of the resulting new clamped OMC design is $g_0/(2\pi)=0.50\ \text{MHz}$.

\label{sec:experiment}
\begin{figure*}[ht]
\centering\includegraphics{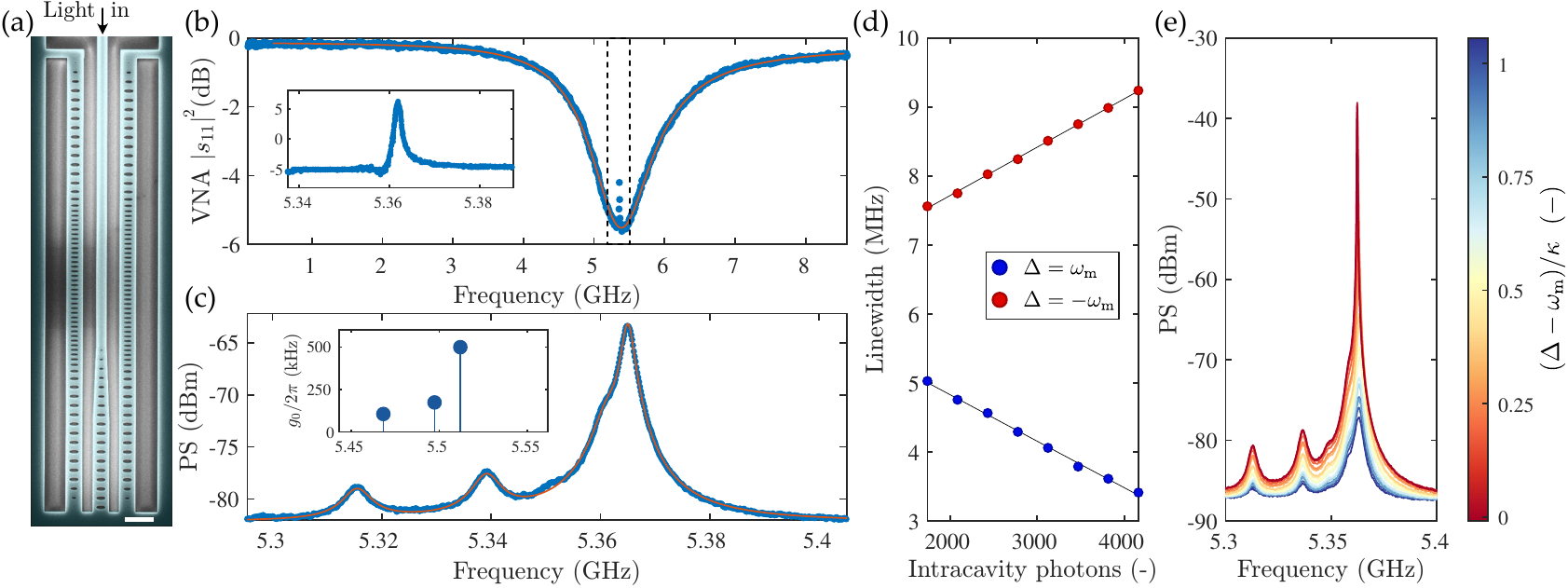}
\caption{\textbf{Fabrication and measurement of the OMC}. \textbf{(a)} Top-down scanning electron micrograph of two silicon clamped OMCs next to an optical bus waveguide. We measure the device in reflection. Scale bar indicates 1 $\mu$m. Light enters through the center waveguide at the top of the figure and couples evanescently to the OMCs. \textbf{(b)} To extract the pump detuning $\Delta$, we perform optical sideband spectroscopy of one of the OMCs measured with a vector network analyzer and a blue-detuned pump laser. The inset shows a zoomed in view of the dashed region with higher resolution where we observe an electromagnetically induced transparency feature at the mechanical frequency. \textbf{(c)} The measured thermal spectrum with model fit (orange) shows a strongly coupled fundamental mechanical mode at $\om/(2\pi) = 5.365$ GHz along with two higher-order modes.
 \textbf{(d)} Mechanical linewidth of the fundamental mechanical mode as a function of optical intracavity photons. We consecutively carry out the experiment with a first blue- then a red-detuned pump. \textbf{(e)} We observe cooperativity of unity and mechanical lasing for an on-chip pump power of $375\ \mu$W.}
\label{fig4}
\end{figure*}

\textbf{Experiments.} We transfer the device pattern to a 220 nm silicon device layer using electron beam lithography (Raith EBPG 5200) followed by an HBr/Cl$_2$-based reactive ion etch (STS ICP MPX). Next, we clean the samples in 3:1 piranha solution before measurement. A top-down scanning electron micrograph of the finished device is shown in Fig.\ref{fig4}a which features two clamped SOI OMCs (false color). We measure the properties of the OMC at room temperature and atmospheric pressure. A pump laser (Santec TSL570) at 1550 nm is injected into an on-chip bus waveguide through focusing grating couplers \cite{benedikovic_high-efficiency_2014} with a coupling efficiency of 18\%. The bus waveguide couples to the OMC evanescently. Light reflected off the OMCs is amplified and subsequently detected on a high-speed photoreceiver (see appendix for details).  

By sweeping the laser wavelength, we detect the optical resonance at $\oo/(2\pi) = 193.1$ THz with a total linewidth $\kappa/(2\pi) = 1.41$ GHz and external coupling rate $\kappa_\mathrm{e}/(2\pi) = 600$ MHz. Next, we modulate the pump intensity and demodulate the reflected signal with a vector network analyzer (Fig.\ref{fig4}b). We thus perform an $s_{11}$ measurement which is used to extract the pump detuning $\Delta = \omega_\mathrm{L} - \oo$ where $\omega_\mathrm{L}$ is the frequency of the pump laser \cite{jasper_laser_2012}. This data is also used to confirm external optical coupling rates. In addition, at an on-chip power level of 375 $\mu$W, we observe an electromagnetically induced transparency window at the mechanical frequency \cite{safavi-naeini_optomechanical_2014, weis_optomechanically_2010}.

Placing the pump blue-detuned from the optical resonance at roughly $\Delta = \om$, we measure the mechanical spectrum in the reflected light with a microwave spectrum analyzer. This reveals several mechanical modes of which we show three in Figure \ref{fig4}c. The full spectrum is shown in appendix.  The spectral spacing and relative optomechanical coupling of the three mechanical modes qualitatively agree with simulation, and the absolute frequencies agree to within 150 MHz (Fig.\ref{fig3}c inset). The fundamental mechanical mode frequency is at $\om/(2\pi) = 5.365$ GHz. Crucially, this puts the device operation in the resolved-sideband regime with $\om/\kappa = 3.6$. To the best of our knowledge, this is the first demonstration of a clamped OMC in this regime. We note a slight asymmetric feature in the fundamental mode (Fig.\ref{fig4}c) which we suspect is related to geometrical disorder. We move on to measure the zero-point optomechanical coupling rate by measuring the mechanical linewidth at varying optical pump powers \cite{aspelmeyer_cavity_2014} for both blue- and red-detuned pumps (Fig.\ref{fig3}d). We find a strong zero-point optomechanical coupling rate of $g_0/(2\pi) = 0.50\pm 0.01$ MHz and a mechanical linewidth of $\gamma/(2\pi) = 6.32$ MHz (Fig.\ref{fig4}d). The zero-point coupling rate is in excellent agreement with simulations. The uncertainty in coupling rate $g_0$ stems mainly from inaccuracy in the measured on-chip powers. The single-photon cooperativity of the device is $\mathcal{C}_\mathrm{0} \equiv 4g_0^2/(\kappa\gamma) = 1.13\cdot10^{-4}$. This cooperativity exceeds that of previously measured clamped devices by about a factor of 7 \cite{sarabalis_release-free_2017,zhang_subwavelength_2022,liu_optomechanical_2022}. We suspect higher single-photon cooperativities are in reach at cryogenic temperatures. To highlight our device performance in relation to state-of-the-art, we present a summary of important parameters for previous clamped work in Table \ref{tab:clamped work}.

\begin{table*}[htbp]
\centering
\caption{\bf Table comparing clamped optomechanical structures on key parameters.}
\begin{tabular}{|c||c|c|c|c|}
\hline
         Reference  & Liu et al. (est.) \cite{liu_optomechanical_2022} & Zhang et al. \cite{zhang_subwavelength_2022} & Sarabalis et al. \cite{sarabalis_release-free_2017}& This work\\
\hline
$g_0/(2\pi)$ (kHz)       & 87 & 51 & 290 & \textbf{500} \\
$\om/(2\pi) $ (GHz)       & \textbf{7.5} & 0.66 & 0.48 & \textbf{5.37} \\
$\kappa/(2\pi) $ (GHz)       & 9.7 & 4.9 & 8.2 & \textbf{1.5} \\
$\gamma/(2\pi) $ (MHz)       & 16 & \textbf{0.6} & 2.6 & 6.3 \\
$\om/\kappa$ (-)       & 0.77 & 0.14 & 0.058 & \textbf{3.6} \\
$\mathcal{C}_\mathrm{0} \equiv 4g_0^2/(\kappa\gamma)$ (-)       & $2.0\cdot10^{-7}$ & $3.5\cdot10^{-6}$ & $1.6\cdot10^{-5}$ & $\mathbf{1.1\cdot10^{-4}}$\\

\hline
\end{tabular}
  \label{tab:clamped work}
\end{table*}

At a pump power of 375 $\mu$W in the bus waveguide with a blue-detuned pump at $\Delta = \omega_{\rm m}$, we reach a cooperativity of unity. This is demonstrated in Fig.\ref{fig4}e where we show mechanical power spectra as the pump detuning is swept closer to $\Delta = \omega_{\rm m}$. When the detuning approaches the mechanical frequency, we observe self-induced oscillations in the fundamental mechanical mode.

In conclusion, we designed and demonstrated a new class of clamped OMCs in SOI leveraging high-wavevector mechanical modes at gigahertz frequencies and counter-propagating optomechanical interactions. To the best of our knowledge, they are the first clamped OMCs in the resolved-sideband regime -- a key requirement for low-noise quantum transduction between optics and mechanics. We observe a record zero-point optomechanical coupling rate for clamped OMCs of $g_0/(2\pi) = 0.50$ MHz -- in excellent agreement with simulation. Their single-photon cooperativity exceeds that of previous clamped OMCs by about an order of magnitude. We suspect that further improvements of the optomechanical overlap are in reach. In addition, clamped OMCs can have significantly larger thermal contact area than suspended structures so they may suffer less from pump-induced mechanical heating in cryogenic environments \cite{meenehan_silicon_2014, ren_two-dimensional_2020}. The operation of our clamped OMCs does not require in-plane bandgaps and relies on robust confinement of mechanical modes with frequencies and wavevectors outside the mechanical continuum. Our approach is not restricted to SOI; it is applicable to a wide range of materials and substrates. The clamped OMCs can be combined with e.g. spins or superconducting qubits. This opens a new avenue for scalable classical and quantum optomechanical circuits for applications in transduction, sensing, and acoustic processing of electromagnetic signals \cite{safavi-naeini_controlling_2019,eggleton_brillouin_2019}. Mechanical systems are often seen as a universal bus. Having them clamped on a substrate while coupling strongly to light unlocks new opportunities in communication and computation.
\newline

\noindent\textbf{Funding.}$\ $\normalsize
We gratefully acknowledge support from the Wallenberg Centre for Quantum Technology and from the European Research Council via Starting Grant 948265.
\newline

\noindent\textbf{Acknowledgement.}$\ $\normalsize
We acknowledge Trond Hjerpekjøn Haug, Witlef Wieczorek, and Per Delsing for helpful discussions. J.K. led the nanofabrication and measurement and assisted with design. P.B. led the design and assisted with nanofabrication and measurement. J.F. assisted with nanofabrication and measurement. J.K., P.B., and R.V.L. wrote the manuscript. R.V.L. provided experimental and theoretical support and conceived as well as supervised the project. 
\newline

\noindent\textbf{Data availability.}$\ $\normalsize
The datasets generated and analyzed for the current study are available from the corresponding author on reasonable request.
\newline

\appendix

\section{Phase-matching for counter-propagating interactions in a standing-wave OMC}
\label{app:phase-match}
As is motivated in the main text, mechanical modes used for optomechanical coupling in clamped OMCs require high wavevectors to be phase-protected from the acoustic continuum. Using mechanical modes with non-zero wavevector implies a varying mechanical phase along the length of the OMC. This is in stark contrast to the $\Gamma$-point mechanical modes commonly adopted in suspended OMCs \cite{chan_optimized_2012,ren_two-dimensional_2020,safavi-naeini_design_2010}. In this section, we explore how this affects the spatial phase of the local optomechanical coupling.  Crucially, we require that there is no significant phase-induced cancellation of coupling when considering a cavity consisting of several unit cells.   

We start by looking at the source of the optomechanical coupling. For the mechanical modes presented in the main text, the moving boundary effect is the most prominent contribution to the coupling. The interaction rate associated with this coupling is \cite{safavi-naeini_design_2010}
\begin{equation}
    g_0 = \sqrt{\frac{\hbar}{2\omega_\mathrm{m}}}\frac{\omega_\mathrm{o}}{2}\frac{\smallint(\Q(\r)\cdot\boldsymbol{n})(\Delta\epsilon|\E^{\parallel}|^2-\Delta(\epsilon^{-1})|\boldsymbol{D}^{\perp}|^2)\d A}{\sqrt{\smallint\rho|\Q(\r)|^2\d^3\r}\smallint\epsilon(\r)|\E(\r)|^2\d^3\r}.
    \label{eq:g0}
\end{equation}
We restrict the following discussion to the contribution of this term; however, the same analysis is valid for the contribution of photoelasticity. Next, we turn our attention to the participating electrical (mechanical) fields $\E$ and $(\Q)$. Crucially, we analyze the coupling between fields in a standing-wave cavity. This implies that the fields consist of both forwards (f) and backwards (b) propagating mechanical and optical components, i.e.,
\begin{equation}
    \begin{alignedat}{2}
        &\E(\r) = (\Ef(\r)e^{ix\kof} + \Eb(\r)e^{ix\kob})f_\mathrm{o}(x)\\
        &\Q(\r) = (\Qf(\r)e^{ix\kmf} + \Qb(\r)e^{ix\kmb})f_\mathrm{m}(x).
    \end{alignedat}
\end{equation}
Here, we assume that the longitudinal direction of the cavity is oriented in the $x$-direction. The fields $\tilde{\E}$ $(\tilde{\Q})$ are unit cell Bloch functions and $f_\mathrm{o}(x)$ $(f_\mathrm{m}(x))$ are envelope functions for the optical (mechanical) fields. For the sake of brevity, we have without loss of generality omitted the presence of pump and sideband optical fields with differing wavevectors. The analysis remains applicable to the full three-wave-mixing scenario involving interactions between a counter-propagating pump and sideband of slightly differing frequencies.

\label{app:exp setup}
\label{sec:experimental setup}
\begin{figure*}[th]
\centering\includegraphics{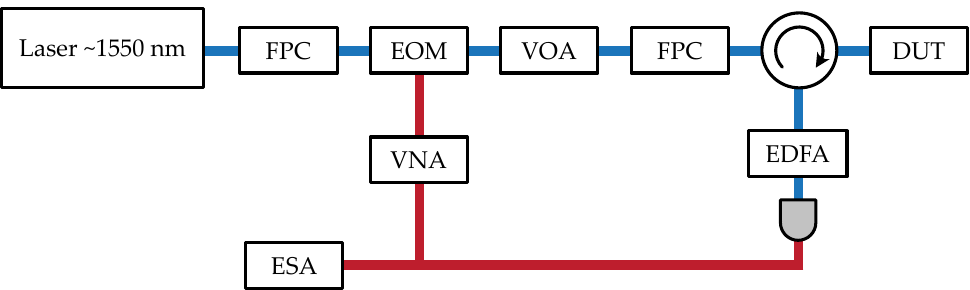}
\caption{Simplified measurement setup used for optical characterization of clamped OMCs. For further info, see Sec. 1\ref{sec:experimental setup}. \textit{Legend:} FPC: Fiber polarization controller.
EOM: (Intensity) Electro-optic modulator.
VOA: Variable optical attenuator.
VNA: Vector network analyzer.
EDFA: Erbium doped fiber amplifier.
ESA: Electrical spectrum analyzer.
DUT: Device under test.
}
\label{figS1}
\end{figure*}
\label{sec:Wide mech spectrum}
\begin{figure*}[t]
\centering\includegraphics{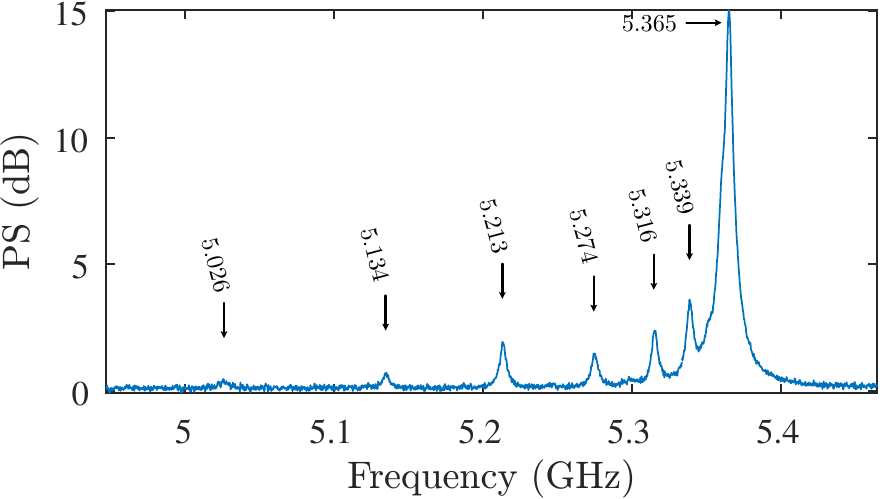}
\caption{Wide range mechanical power spectrum. We subtract the background by using data taken with a far blue-detuned pump, i.e. $\Delta \gg \om$.  In addition to the three modes presented in the main text, we detect four more mechanical modes.}
\label{figS3}
\end{figure*}

We can now use \eqref{eq:g0} to calculate the total coupling. First, we assume that the envelopes varies slowly, i.e. the cavity consists of many unit cells, we can therefore take the envelope functions to be effectively uniform along the OMC to illustrate the principles at work. Due to the squared optical fields, the result can be expanded into six terms with differently varying phase along the $x$-direction. The terms have factors on the following forms
\begin{equation}
    \begin{alignedat}{2}
        &\smallint e^{ix(\kmf)}(...) \d A,\qquad\qquad &(i)\\
        &\smallint e^{ix(\kmf+\kof -\kob)}(...) \d A,\qquad\qquad &(ii)\\
        &\smallint e^{ix(\kmf-\kof +\kob)}(...) \d A,\qquad\qquad &(iii)\\
        &\smallint e^{ix(\kmb)}(...) \d A,\qquad\qquad &(iv)\\
        &\smallint e^{ix(\kmb+\kof -\kob)}(...) \d A,\qquad\qquad &(v)\\
        &\smallint e^{ix(\kmb-\kof +\kob)}(...) \d A.\qquad\qquad &(vi)\\
    \end{alignedat}
    \label{eq:terms}
\end{equation}
In the case that the mechanics has vanishing wavevector like in most suspended OMCs we have $\kmf = \kmb = 0$ such that the terms $(i)$ and $(iv)$ will lose their spatial phase, preventing cancellations in the total overlap integral. For the remaining terms to generate finite coupling, we find that $\kof = \kob$ in this scenario with small mechanical wavevector. This however leads to a contradiction that standing waves requires counter-propagating fields. Therefore, in a sufficiently long cavity, only two terms will contribute to the overall coupling and other terms are strongly suppressed. In essence, we see that low-wavevector mechanics couples well to co-propagating optical pump and sideband as is familiar from forward intra-modal Brillouin interactions \cite{eggleton_brillouin_2022}.

On the other hand, in our clamped OMCs we have $\kmf = -\kmb = k_\mathrm{m} \neq 0$ such that the integrands of $(i)$ and $(iv)$ tend to be strongly suppressed due to cancellations arising from different parts of the OMC. The phase-matching condition therefore becomes $k_\mathrm{m} = \pm(\kof - \kob),$ where the plus (minus) sign satisfies coupling in terms $(iii)$ and $(v)$ ($(ii)$ and $(vi)$). As an example, when $\kof = -\kob = k_\mathrm{o}$ we have that $k_\mathrm{m} = 2k_\mathrm{o}$ which is the principle upon which coupling is generated in the OMCs presented in the main text. This condition is familiar from backward intra-modal Brillouin interactions \cite{eggleton_brillouin_2022}.

Generally, integration over terms in \eqref{eq:terms} with phase mismatch $\Delta k$ leads to suppression by a factor $\sin(\Delta k L)/\Delta k L$. As expected, we see that the implications of phase-matching are most prominent for long cavities where wavevectors are well-defined. For the OMC presented in the main text, co-propagating terms involving our high-wavevector mechanical mode are suppressed by a factor $\Delta k L = 2\ko L \approx \pi N \approx 10^{2}$ compared to counter-propagating terms. Here we used $\ko \approx \pi/(2a)$ and $L=Na$ with $N=31$ the number of unit cells. 

In summary, this analysis shows that phase-matched counter-propagating optomechanical interactions are possible in OMCs when using high-wavevector mechanical modes. Therefore, these interactions can be as strong and scale similarly to the more common approach based on co-propagating optomechanical interactions and low-wavevector mechanical modes.

\section{Experimental setup}

A simplified diagram of the measurement setup is presented in Fig.\ref{figS1}. We carry out the optical characterization with a fiber-coupled continuously tunable laser in the C-band (Santec TSL570). An electro-optic intensity modulator (IXblue MX-LN-20) generates sideband tones at the mechanical frequency. To control the power sent to the device under test, we use digitally controlled variable optical attenuators (Sercalo VP1-9N-12-16). Light polarization in the fiber network is managed through fiber polarization controllers.  

After interacting with the device under test, the reflected light is circulated and amplified with an erbium doped fiber amplifier (Amonics AEDFA-PA-35-B). A high-speed photoreceiver (Newport 1544-B) downconverts the GHz signals contained in the reflected light. We detect and display the thermal mechanical power spectrum with an electrical spectrum analyzer (R\&S  FSW26) in real-time mode with a resolution bandwidth of 250 kHz. Finally, we use a vector network analyzer (R\&S ZNB20) to drive and demodulate the GHz optical modulation. 

\section{Wide mechanical spectrum}

The mechanical spectrum shown in Fig.\ref{fig4}c in the main text features three prominent mechanical modes. Searching in a wider span of frequencies reveals four additional modes. We present the spectrum along with the measured frequencies in Fig. \ref{figS3}. To reduce the risk of misinterpreting laser noise as mechanical modes, we perform the measurement with two independent laser sources (Santec TSL570 and Toptica CTL1550).

\bibliography{clampedOMC}

\end{document}